\documentclass[aps,twocolumn]{revtex4}

\usepackage{graphicx}
\usepackage{color}
\usepackage{multirow}
\usepackage{comment}

\usepackage{ulem}
\begin{document}

\bibliographystyle{apsrev}

\title{Astrophysical Model Selection in Gravitational Wave Astronomy}
\author{\surname {Matthew} R. Adams and {Neil} J. Cornish}
\affiliation{Department of Physics, Montana State University, Bozeman, MT 59717}

\author{\surname {Tyson} B. Littenberg}\affiliation{Maryland Center
  for Fundamental Physics, Department of Physics, University of
  Maryland, College Park, MD 20742}\affiliation{Gravitational
  Astrophysics Laboratory, NASA Goddard Spaceflight Center, 8800
  Greenbelt Rd., Greenbelt, MD 20771}

\date{\today}

\begin{abstract}
Theoretical studies in gravitational wave astronomy have mostly focused on the information that
can be extracted from individual detections, such as the mass of a binary system and its location in space.
Here we consider how the information from multiple detections can be used to constrain astrophysical
population models. This seemingly simple problem is made challenging by the high dimensionality and
high degree of correlation in the parameter spaces that describe the signals, and by the complexity of the
astrophysical models, which can also depend on a large number of parameters, some of which might not be
directly constrained by the observations. We present a method for constraining population models using a
Hierarchical Bayesian modeling approach which simultaneously infers the source parameters and population
model and provides the joint probability distributions for both. We illustrate this approach by considering
the constraints that can be placed on population models for galactic white dwarf binaries using a future
space based gravitational wave detector. We find that a mission that is able to resolve $\sim 5000$ of the
shortest period binaries will be able to constrain the population model parameters, including the chirp mass
distribution and a characteristic galaxy disk radius to within a few percent. This compares favorably to
existing bounds, where electromagnetic observations of stars in the galaxy constrain disk radii to
within $20\%$. 
\end{abstract}

\pacs{}
\maketitle

\section{Introduction}
There is an old joke in astrophysics that with one source you have a discovery, and with two you
have a population. With a population of sources it becomes possible to constrain astrophysical models.
Until recently, studies of milli-Hertz gravitational wave science have either focused on making predictions
about the source populations, or have looked at detection and parameter estimation for individual sources.
These types of studies have featured heavily in the science assessment of alternative space-based
gravitational wave mission concepts, where metrics such as detection numbers and histograms of the parameter
resolution capabilities for fiducial population models were used to rate science performance
(see {\it eg.} Ref.~\cite{2009CQGra..26i4014S}). These are certainly useful metrics, but they only
tell part of the story. A more powerful and informative measure of the science capabilities is the ability
to discriminate between alternative population models.

Inferring the underlying population model, and the attendant astrophysical processes responsible for the
observed source distribution, from the time series of a gravitational wave detector is the central
science challenge for a future space mission. It folds together the difficult task of identifying
and disentangling the multiple overlapping signals that are in the data, inferring the individual source
parameters, and reconstructing the true population distributions from incomplete and imperfect information. 

The past few years have seen the first studies of the astrophysical model selection problem in the
context of space based gravitational astronomy. Gair and
collaborators~\cite{Gair:2010yu,Gair:2010bx,Sesana:2010wy,AmaroSeoane:2012je} have looked at
how extreme mass ratio inspiral (EMRI) formation
scenarios and massive black hole binary assembly scenarios can be constrained by GW observations using
Bayesian model selection with a Poisson likelihood function. Plowman and
collaborators~\cite{Plowman:2009rp, Plowman:2010fc} have performed
similar studies of black hole population models using a frequentist approach based on error kernels and
the Kolmogorov-Smirnov test. Related work on astrophysical model selection for ground based
detectors can be found in Refs.~\cite{Mandel:2009pc, O'Shaughnessy:2012zc}.

We develop a simple yet comprehensive Hierarchical Bayesian modeling approach that uses the full
multi-dimensional and highly correlated parameter uncertainties of a collection of signals to constrain
the joint parameter distributions of the underlying astrophysical models.  The method is general and can
be applied to any number of astrophysical model selection
problems~\cite{2009ApJ...704..629M, 2012arXiv1206.3540S, 2012arXiv1208.3036L}

A remarkable feature of the Hierachial Bayesian method is that in its purest form it is completely free
of selection effects such as Malmquist bias. By ``purest form'' we mean where the signal model
extends over the entire source population, including those with vanishingly small
signal-to-noise ratio~\cite{Messenger:2012jy}. In practice it is unclear how to include arbitrarily
weak sources in the analysis, and in any case the computational cost would be prohibitive, so
we are forced to make some kind of selection cuts on the signals, and this will introduce a bias if
left uncorrected~\cite{Schutz:2011tw}. 

To illustrate the Hierachical Bayesian approach and to investigate where bias can arise, we look at the problem of
determining the population model for white dwarf binaries in the Milky Way.  Future space based missions
are expected to detect thousands to tens of thousands of white dwarf binaries~\cite{AmaroSeoane:2012je, Crowder:2006eu, Nissanke:2012eh, Timpano:2005gm, Littenberg:2011zg}. Here we focus on determining the spatial
distribution and the chirp mass distribution, but in future work we plan to extend our study to include
a wider class of population characteristics such as those described in Ref.~\cite{Nissanke:2012eh}.
Determining the galaxy shape using gravitational wave observations of white dwarf binaries will be an
independent measure on the shape of the galaxy to complement electromagnetic observations.  Additionally,
the white dwarf binaries that are not detectable form a very bright stochastic foreground.  Accurately
modeling the confusion foreground level is crucial for the detection of extragalactic stochastic
gravitational wave signals~\cite{2010PhRvD..82b2002A}.

The paper is organized as follows: The Hierarchical Bayesian approach is described in \S~\ref{HB}, and is
illustrated using a simple toy model in \S~\ref{toy1}. A more realistic toy model is developed in
\S~\ref{toy2} to explore mis-modeling biases that can occur when using Gaussian approximations to
the likelihood function. In \S~\ref{galaxy} the method is applied to simulated observations of galactic
white dwarf binaries, and in \S~\ref{approx} the possibility of using the Fisher Information Matrix
approximation to the likelihood is explored. Concluding thoughts follow in \S~\ref{concl}.

\section{Hierarchical Bayesian Modeling}\label{HB}

Hierarchical Bayesian modeling has been around since at least the
1950's~\cite{Good:1965,1972_LindleySmith,Morris:1992,MacKay94},
but it is only now becoming widely known and used.  The term ``hierarchical" arises because the analysis has
two levels. At the highest level are the space of models being considered, and at the lower level are the
parameters of the models themselves. Hierachical Bayes provides a method to simultaneously perform model
selection and parameter estimation. In this work we will consider models of fixed dimension that
can be parameterized by smooth functions of one or more hyper-parameters. The joint posterior
distribution for the model parameters $\vec{\lambda}$ and the hyper-parameters $\vec{\alpha}$ given
data $s$ follows from Bayes' theorem:
\begin{equation}\label{formal}
p(\vec{\lambda}, \vec{\alpha} \vert s) = \frac{ p(s \vert \vec{\lambda}, \vec{\alpha}) p(\vec{\lambda}
 \vert \vec{\alpha}) p(\vec{\alpha})}{ p(s)} \, ,
\end{equation}
where $p(s \vert \vec{\lambda}, \vec{\alpha})$ is the likelihood, $p(\vec{\lambda} \vert \vec{\alpha})$ is
the prior on the model parameters for a model described by hyper-parameters $\vec{\alpha}$,
$p(\vec{\alpha})$ is the hyper-prior and $p(s)$ is a normalizing factor
\begin{equation}
p(s) = \int p(s, \vec{\alpha}) d\vec{\alpha}= \int p(s \vert \vec{\lambda}, \vec{\alpha}) p(\vec{\lambda}
 \vert \vec{\alpha}) p(\vec{\alpha})  d\vec{\lambda} d\vec{\alpha} \, .
\end{equation}
The quantity $p(s, \vec{\alpha})$ can be interpreted as the ``density of evidence'' for a model with
hyper-parameters $\vec{\alpha}$.

The integral marginalizing over the hyper-parameters is often only tractable numerically, and this can be
computationally expensive. Empirical Bayes is a collection of methods that seek to estimate the hyper-parameters
in various ways from the data~\cite{Casella:1985, Carlin:2000}. Markov chain Monte Carlo (MCMC) techniques
allow us to implement Hierachical Bayesian modeling without approximation by producing samples from the
joint posterior distributions, which simultaneously informs us about the model parameters $\vec{\lambda}$ and the
hyper-parameters $\vec{\alpha}$. This approach helps reduce systematic errors due to mis-modeling, as the data
helps select the appropriate model. An example of this is the use of hyper-parameters in the instrument noise
model, such that the noise spectral density is treated as an unknown to be determined from the
data~\cite{Cornish:2007if, Littenberg:2010gf, 2010PhRvD..82b2002A}. 

Hierarchical Bayesian modeling can be extended to discrete and even disjoint model spaces using the
Reverse Jump Markov Chain Monte Carlo (RJMCMC)~\cite{green_highly_2003} algorithm. Each discrete
models can be assigned its own set of continuous hyper-parameters.

\section{Toy Model I}\label{toy1}

As a simple illustration of hierarchical Bayesian modeling, consider some population of $N$ signals,
each described by a single parameter $x_i$ that is drawn from a normal distribution with standard
deviation $\alpha_0$. The measured values of these parameters are affected by instrument noise that
is drawn from a normal distribution with standard deviation $\beta$. The maximum likelihood
value for the parameters is then $\bar{x}_i = \alpha_0 \delta_1 + \beta \delta_2$ where the $\delta$'s are
i.i.d. unit standard deviates. Now suppose that we employ a population model where the parameters are
distributed according to a normal distribution with standard deviation $\alpha$. Each choice
of $\alpha$ corresponds to a particular model with posterior distribution
\begin{equation}\label{eq:toy1post}
p(x_i\vert s, \alpha) = \frac{1}{p(s,\alpha)} \prod_{i=1}^N \frac{1}{(2\pi \alpha\beta)} 
e^{-(\bar{ x}_i - x_i)^2/2\beta^2} e^{-x_i^2/2\alpha^2} \, ,
\end{equation}
and model evidence
\begin{equation}\label{mla}
p(s,\alpha) = \frac{1}{(\sqrt{2\pi}\sqrt{\alpha^2+\beta^2})^N}\prod_i e^{-{\bar x}_i^2 /2 (\alpha^2+\beta^2)} \, .
\end{equation}
To arrive at a Hierarchical Bayesian model we elevate $\alpha$ to a hyper-parameter and introduce a hyper-prior
$p(\alpha)$ which yields the joint posterior distribution
\begin{equation}\label{eq:toy1post2}
p(x_i, \alpha \vert s) = \frac{p(x_i\vert s, \alpha) p(\alpha)}{p(s)} \, .
\end{equation}
Rather than selecting a single ``best fit'' model, Hierarchical Bayesian methods reveal the range of models that are
consistent with the data. In the more familiar, non-hierarchical approach we would maximize the model
evidence (\ref{mla}) to find the model that best describes the data, which is here given by
\begin{equation}
\alpha_{\rm ME}^2 = \frac{1}{N}\sum_{i=1}^N {\bar x}_i^2 - \beta^2 ,
\end{equation}
Since ${\rm Var}({\bar x}_i)=\alpha_0^2+\beta^2$, we have
\begin{equation}\label{estx}
\alpha_{\rm ME}^2 = \alpha_0^2 \pm {\cal O}(\sqrt{2}(\alpha_0^2+\beta^2)/\sqrt{N}) \, .
\end{equation}
The error estimate comes from the sample variance of the variance estimate. In the limit that the
experimental errors $\beta$ are small compared to the width of the prior $\alpha_0$, the error in
$\alpha$ scales uniformly as $1/\sqrt{N}$. The scaling is more complicated when we have a collection
of observations with a range of measurement errors. Suppose that the measurement errors are large compared to
the width of the prior, and that we have $N_1$ observations with standard error $\beta_1$, $N_2$ observations
with standard error $\beta_2$ {\it etc}, then the error in the estimate for $\alpha$ is
\begin{equation}
\Delta \alpha^2 = \left(\sum_i \frac{N_i}{\beta_i^4}\right)^{-1/2} \, .
\end{equation}
Recalling that $1/\beta_i$ scales with the signal-to-noise ratio of the observation, we see
that a few high SNR observations constrain $\alpha$ far more effectively than a large number of
low SNR observations.

The above calculation shows that the maximum evidence criteria provides an unbiased estimator for
the model parameter $\alpha_0$, but only if the measurement noise is 
consistently included in both the likelihood and the simulation
of the $\bar{x}_i$. Using the likelihood from (\ref{eq:toy1post}) but
failing to include the noise in the simulations leads to the biased estimate
$\alpha^2_{\rm ME} = \alpha_0^2 - \beta^2$. Conversely, including noise in the simulation and
failing to account for it in the likelihood leads to the biased estimate 
$\alpha^2_{\rm ME} = \alpha_0^2 + \beta^2$.
These same conclusions apply to the Hierarchical Bayesian approach,
as we shall see shortly.

\subsection{Numerical Simulation}

\begin{figure}[htbp]
   \centering
   \includegraphics[width=3.0in,angle=0] {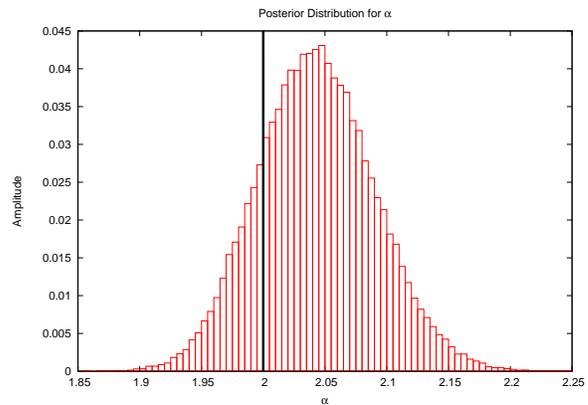} 
   \caption{The Marginalized Posterior Distribution Function for $\alpha$. The injected value is indicated by
the vertical black line.}
   \label{fig:toy1_alpha}
\end{figure}

The joint posterior distribution (\ref{eq:toy1post2}) can be explored using MCMC techniques. To do this
we produced simulated data with $N= 1000$, $\alpha_0 = 2$ and $\beta=0.4$ and adopted a flat hyper-prior
for $\alpha$. The posterior distribution function for $\alpha$, marginalized over the $x_i$, is shown
in Figure~\ref{fig:toy1_alpha}. The distribution includes
the injected value, and has a spread consistent with the error estimate of (\ref{estx}). The
Maximum-a-Posteriori (MAP)
estimate for $\alpha$ has been displaced from the injected value of $\alpha_0 = 2$ by the simulated noise.

\begin{center}
\begin{figure}[htbp]
   \centering
   \includegraphics[width=3.0in] {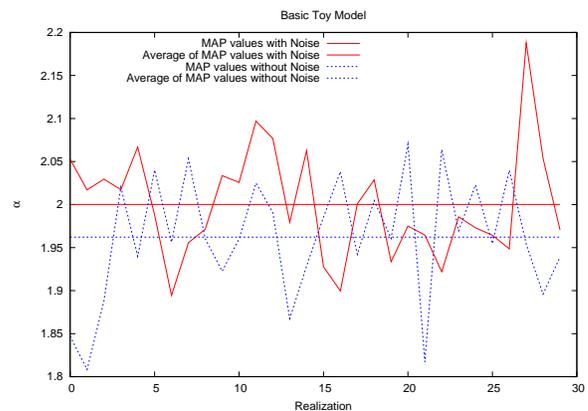} 
   \caption{MAP values for 30 different simulations of the toy model.  The red curve includes noise in the
simulated signal and converges to $\alpha_0$ as expected.  The blue curves does not include noise in the
simulation and converges to $\alpha_0^2 - \beta^2$}
   \label{fig:toy1_MAPs}
\end{figure}
\end{center}

To test that there is no bias in the recovery of the model hyper-parameter $\alpha$, we produced 30 different
realizations of the data and computed the average MAP value.  Figure~\ref{fig:toy1_MAPs} shows the MAP value
for each of these realizations and the corresponding average. We see that as we average over
multiple realizations $\alpha$ does indeed converge to the injected value. The blue line in
Fig.~\ref{fig:toy1_MAPs} shows a biased recovery for $\alpha$ when noise is not included in the data. 
We instead recover $\alpha = \sqrt{\alpha^2_0 - \beta^2} \approx 1.96$.

\section{Toy Model II}\label{toy2}

The Hierarchical Bayesian approach produces un-biased estimates for the model parameters if the signal and
the noise (and hence the likelihood) are correctly modeled. However, in some situations the
cost of computing the likelihood can be prohibitive, and it becomes desirable to use
approximations to the likelihood, such as the Fisher Information Matrix. For example, to
investigate how the design of a detector influences its ability to discriminate between different astrophysical
models, it is necessary to Monte Carlo the analysis over many realizations of the source population for many
different instrument designs, which can be very costly using the full likelihood.

To explore these issues we introduce a new toy model that more closely resembles the likelihood functions
encountered in gravitational wave data analysis. Consider a waveform $h_0$ that represents a single data
point ({\it e.g.} the amplitude of
a wavelet or a Fourier component), which can be parameterized in terms of the distance to the source $d_0$. The
instrument noise $n$ is assumed to be Gaussian with variance $\beta^2$. Here we will treat the noise level
$\beta$ as a hyper-parameter to be determined from the observations. Adopting a fiducial noise level $\beta_0$
allows us to define a reference signal-to-noise ratio ${\rm SNR}^2_0 = h_0^2/\beta_0^2$.
The likelihood of observing data $s= h_0+n$ for a source at distance $d$ with noise level $\beta$ is then
\begin{equation}\label{liketoy}
p(s\vert d, \beta)  = \frac{1}{\sqrt{2\pi}\beta}e^{-(s-h)^2/(2\beta^2)}
\end{equation}
where $h = (d_0/d) h_0$. The likelihood is normally distrubuted in the inverse distance $1/d$, with a maximum
that depends on the particular noise realization $n$:
\begin{equation}
\frac{1}{d_{\rm ML}} = \frac{1+n/(\beta_0 {\rm SNR}_0)}{d_0} \, .
\end{equation}
Now suppose that the distances follow a one-sided normal distribution
$p(d \geq 0) = \frac{2}{\sqrt{2\pi}\beta}\exp(-d^2/2\alpha_0^2)$, and that we adopt a corresponding model
for the distance distribution with hyper-parameter $\alpha$ and a flat hyper-prior.

\begin{figure}[htbp]
   \centering
   \includegraphics[width=3.0in,angle=0] {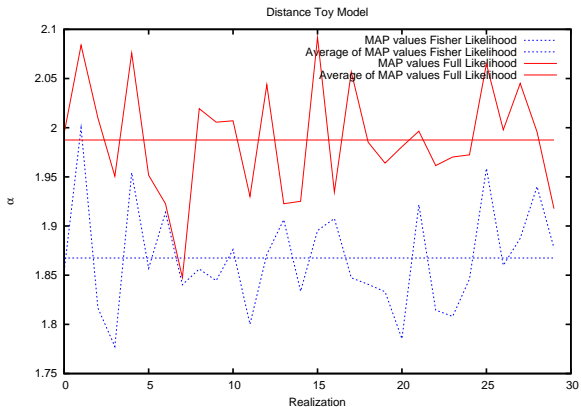} 
   \caption{MAP values for 30 different realizations of the toy model II. Using the full likelihood (blue) the
MAP values converge to the injected value, but with the Fisher Matrix approximation to the likelihood (red) there
is a bias.}
   \label{fig:toy2_MAPs}
\end{figure}

We simulate the data from $N=1000$ sources with $\alpha_0=2$ and $\beta = 0.05$. The values of $\alpha_0$
and $\beta$ were chosen to give a fiducial ${\rm SNR}=5$ for $d = 2\alpha_0$. In the first of our
simulations the value of $\beta$ was assumed to be known and we computed the MAP estimates of
$\alpha$ for 30 different simulated data sets. As shown in Figure~\ref{fig:toy2_MAPs}, the
average MAP estimate for $\alpha$ converges to the injected value.

\begin{figure}[htbp]
   \centering
   \includegraphics[width=3.0in,angle=0] {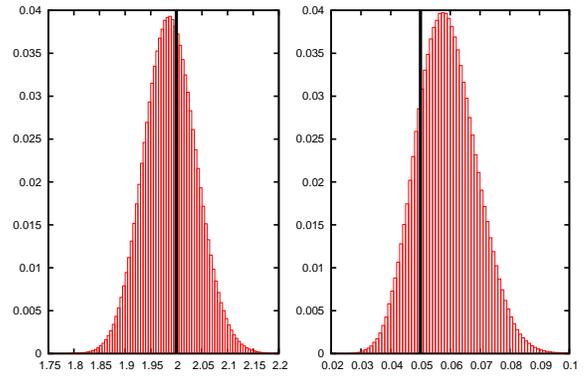} 
   \caption{PDFs for the prior hyper-parameter $\alpha$ and the noise level $\beta$ for toy model II.  Both are
individually constrained in this model.  The injected values are shown by the black lines.}
   \label{fig:BetaAlpha}
\end{figure}

In contrast to the first toy model where only the combination $\alpha^2+\beta^2$ is
constrained by the data, in this more realistic toy model both the noise level $\beta$ and
the model hyper-parameter $\alpha$ are separately constrained.  Figure~\ref{fig:BetaAlpha}
shows the marginalized PDFs for both $\beta$ and $\alpha$. Tests using
multiple realizations of the data show that the MAP values of $\alpha$ and $\beta$
are un-biased estimators of the injected parameter values.

\subsection{Approximating the Likelihood}

For stationary and Gaussian instrument noise the log likelihood for a signal described by
parameters $\vec{\lambda}$ is given by
\begin{equation}\label{logL}
L(\vec{\lambda}) =  -\frac{1}{2}(s-h(\vec{\lambda}) \vert s-h(\vec{\lambda}))
\end{equation}
where $(a\vert b)$ denotes the standard noise-weighted inner product, and we have
supressed terms that depend on
the noise hyper-parameters. We can expand the waveform
$h(\vec{\lambda})$ about the injected source parameters $\vec{\lambda}_0$:
 \begin{equation}
h(\vec{\lambda}) =  h(\vec{\lambda}_0) + \Delta \lambda^i \bar{h}_{,i} + \Delta \lambda^i \Delta \lambda^j \bar{h}_{,ij}
+ \mathcal{O}(\Delta \lambda^3)
\end{equation}
where $\Delta \vec{\lambda} = \vec{\lambda}-\vec{\lambda}_0$, and it is understood that the derivatives are
evaluated at $\vec{\lambda}_0$. Expanding the log likelihood we find:
\begin{eqnarray}\label{lex}
L(\Delta \vec{\lambda}) =&-&\frac{1}{2} (n\vert n) +\Delta \lambda^i (n \vert h_{,i})\nonumber \\
&-&\frac{1}{2} \Delta \lambda^i \Delta \lambda^j (h_{,i}\vert h_{,j})+ {\cal O}(\Delta \lambda^3)\, .
\end{eqnarray}
The maximum likelihood solution is found from $\partial L/\partial\Delta \lambda^i = 0$, which yields
$\Delta \lambda_{\rm ML}^i = (n \vert h_{j}) \Gamma^{ij}$, where $\Gamma^{ij}$ is the inverse of the Fisher
Information Matrix $\Gamma_{ij}=(h_{,i}\vert h_{,j})$. Using this solution to
eliminate $(n \vert h_{,i})$ from (\ref{lex}) yields the quadratic, Fisher Information
Matrix approximation to the likelihood:
\begin{equation}\label{fish}
L(\vec{\lambda}) = {\rm const.} -\frac{1}{2} (\lambda^i - \lambda^i_{\rm ML})(\lambda^j- \lambda^j_{\rm ML}) \Gamma_{ij} \; .
\end{equation}
This form of the likelihood can be used in simulations by drawing the $\Delta \lambda_{\rm ML}^i$ from a
multi-variate normal distribution with covariance matrix $\Gamma^{ij}$.

In our toy model $\Gamma_{dd} = {\rm SNR}^2_0 \beta_0^2/(\beta^2 d_0^2)$, and
$L(d)=-{\rm SNR}_0^2 \beta_0^2 (d-d_{\rm ML})^2/(2 \beta^2 d_0^2)$. The approximate likelihood
follows a normal distribution in $d$ while the full likelihood follows a normal distribution
in $1/d$. For signals with large SNR this makes little difference, but at low SNR the difference
becomes significant and results in a bias in the recovery of the model hyper-parameters, as shown in
Figure~\ref{fig:toy2_MAPs}. In this instance there is a simple remedy: using $u = 1/d$ in place of
$d$ in the quadratic approximation to the likelihood exactly reproduces the full likelihood in this
simple toy model. However, it is not always so easy to correct the deficiencies
of the quadratic, Fisher Information Matrix approximation to the likelihood.

\section{White Dwarf Binaries in the Milky Way}\label{galaxy}

To illustrate how the Hierarchical Bayesian approach can be applied to an astrophysically relevant
problem, we investigate how population models for the distribution of white dwarf binaries in
the Milky Way galaxy can be constrained by data from a space based gravitational wave detector.
Several studies have looked at
parameter estimation for individual white dwarf binaries in the Milky
Way~\cite{Cornish:2003vj, Vecchio:2004ec, babak:WDs}.  We extend these studies to consider how the
individual observations can be combined to infer the spatial and mass distributions of white
dwarf binaries in the Galaxy.

We use the Laser Interferometer Space Antenna (LISA)~\cite{LISAwhitepaper} as our reference mission.
We focus this analysis on short-period galactic binaries, with gravitational wave frequencies above 4 mHz.
Our conclusions would be little changed if we considered the recently proposed eLISA~\cite{AmaroSeoane:2012je}
mission instead, as both are able to detect roughly the same number of galactic binaries in the frequency bands 
considered here.

\begin{figure}[htbp]
   \centering
   \includegraphics[width=3.0in,angle=0] {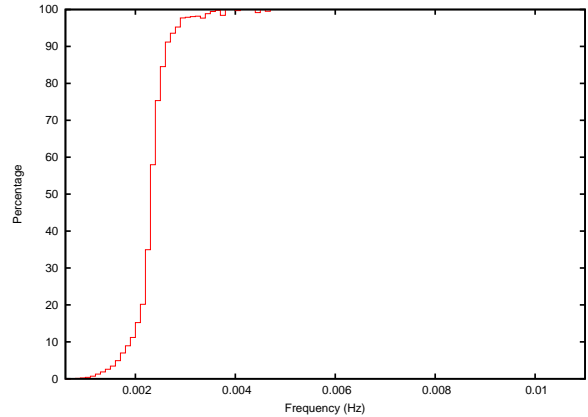} 
   \caption{ The percentage of sources which are detectable as a function of frequency.  Virtually 100\% of the white dwarf binaries in the Milky
   Way above 4mHz would be detected by LISA.}
   \label{fig:CumFreq}
\end{figure}

The 4 mHz lower limit is chosen to simplify the analysis in two ways. Firstly, it avoids the signal overlap
and source confusion problems that become significant at lower frequencies~\cite{Crowder:2006eu}, and secondly,
it circumvents the issue of sample completeness and Malmquist selection bias since LISA's coverage of the
galaxy is complete at high frequencies.   This claim is substantiated in Figure~\ref{fig:CumFreq} showing the cumulative percentage of binaries detected as a function of frequency for
a 4 year LISA mission. A given frequency bin represents the percentage of binaries of that frequency and
higher that are detected. All binaries above $\sim4$ mHz are detectable by LISA, of which there are $\sim 5000$.  

It is possible to extend our analysis to include all detectable white dwarf
binaries if we were to properly account for the undetectable sources. One way to do this is to convolve
the astrophysical model priors by a function that accounts for the selection effects~\cite{Schutz:2011tw} so
that we are working with the predicted observed distribution rather than the theoretical distribution.
Another approach is to marginalize over the un-detectable signals~\cite{Messenger:2012jy}.

The high frequency signals are not only the simplest to analyze, but they  also tend to have the highest
signal-to-noise ratios, the best sky localization, and the best mass and distance determination due to
their more pronounced evolution in frequency. When simulating the population of detectable sources we will assume that binaries of all frequencies above 
4 mHz
are homogeneously distributed throughout the galaxy and share the same chirp mass distribution. In reality
the population is likely to be more heterogenous, and more complicated population models will have to be
used.

\subsection{Likelihood}

The likelihood for a single source is given by:
\begin{equation}\label{eq:FullLikelihood}
p(s \vert \vec{\lambda}) = C e^{-\left(s - h(\vec{\lambda}) | s - h(\vec{\lambda})\right)/2} \, .
\end{equation}
Here $p(s \vert \vec{\lambda})$ is the likelihood that the residual $s - h(\vec{\lambda})$ is
drawn from Gaussian noise, where $s$ is the data, and $h(\vec{\lambda})$ is the signal produced in the detector
by a source described by parameters $\vec{\lambda}$.   The simulated data $s=h(\vec{\lambda}_0)+n$ includes
a waveform $h(\vec{\lambda}_0)$ and a realization of the LISA instrument noise $n$.  The
normalization constant $C$ depends on
the instrument noise levels, but is independent of the waveform parameters.  

The waveform for a white dwarf binary is well approximated by:
\begin{eqnarray}
h_+(t) &=& \frac{1}{d}\frac{4G\mathcal{M} \Omega^2}{c^4}\left(\frac{1+\cos^2\iota}{2}\right)\cos(\Omega t) \nonumber \\
h_{\times}(t) &=& \frac{1}{d}\frac{4G\mathcal{M}\Omega^2}{c^4}\cos{\iota}\sin(\Omega t)
\end{eqnarray}
where $\Omega=2\pi f$.  We have 8 parameters that describe a white dwarf binary signal, the frequency $f$,
the distance to the source $d$, the chirp mass $\mathcal{M}$, the inclination angle $\iota$,
a polarization angle $\psi$, a phase angle $\varphi_0$, and sky location parameters $\theta$ and $\phi$.
To leading order, the frequency evolves as:
\begin{equation}\label{fdot}
\dot{f} = \frac{96\pi}{5} (\pi \mathcal{M})^{5/3} f^{11/3} \, .
\end{equation}
Sources with $\dot{f} \, T^2 \, {\rm SNR} \sim 1$, where $T$ is the observation time, provided useful measurments
of the chirp mass $\mathcal{M}$ and the distance $d$~\cite{Schutz:1986gp, Takahashi:2002ky}.
The strong $f$ dependence in (\ref{fdot}) is the reason why high frequency binaries are the best candidates for 
placing strong constraints on the distance and chirp mass.

\subsection{Model for the Galaxy}

We adopt a bulge plus disk model for the galaxy shape~\cite{Nelemans:2003ha, Nelemans:2000es, Nelemans:2001nr,
Nelemans:2001hp}. Choosing the x-y plane as the plane of the galaxy, the density of stars in the galaxy is given by:
\begin{equation}
\rho(x, y, z) = \rho_0\left(A e^{-r^2/R_b^2} + (1-A)e^{-u/R_d} \text{sech}^2{(z/Z_d)}\right)
\end{equation}
Here, $r^2 = x^2+y^2+z^2$, $u^2=x^2+y^2$, $R_b$ is the characteristic radius for the bulge, and $R_d$ and $Z_d$
are a characteristic radius and height for the disk respectively.   The quantity $\rho_0$ is a reference density
of stars and the coefficient $A$, which ranges between 0 and 1, weights the number of stars in the bulge versus
the number in the disk.  We produced synthetic galaxies using the catalog of binaries provided by Gijs Nelemans
for the Mock LISA Data Challenges (MLDC)~\cite{Arnaud:2007jy}.

With appropriate normalization, the spatial density $\rho$ becomes our prior distribution for the spatial
distribution of galactic binaries. The parameters of the density distribution $A,$ $R_b,$ $R_d$ and $Z_d$ become
hyper-parameters in the Hierarchical Bayesian analysis. Each set of values for the four parameters corresponds
to a distinct model for the shape of the galaxy.  For our simulations, we chose a galaxy with $A=0.25$,
$R_b=500$pc, $R_d=2500$pc, and $Z_d=200$pc.

\subsection{Chirp Mass Prior}

Our ability to measure the hyper-parameters of the spatial distribution depends on how well we measure
the sky location and distance for each binary.  For many sources, the distance is
poorly determined because it is highly correlated with the chirp mass. However, there are enough binaries
that have sufficiently high frequency, chirp mass and/or SNR to provide tight constraints
on the chip mass distribution.  The empirically determined chirp mass distribution then functions as a
prior for the lower SNR, less massive, or lower frequency sources, and improves their distance determination.

\begin{figure}[htbp]
   \centering
   \includegraphics[width=3.0in,angle=0] {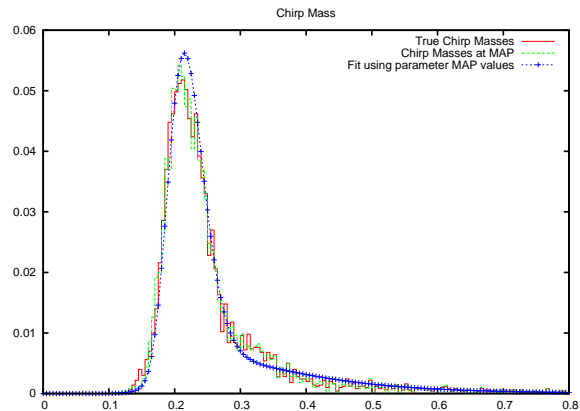} 
   \caption{The chirp mass distribution of the 5000 binaries used in our simulations is shown in red.
The green distribution shows the MAP values of the recovered chirp mass for each binary, and the blue shows
the model (\ref{ChirpDist}) using the MAP values for the chirp mass prior hyper-parameters. The brightest
binaries accurately capture the chirp mass distribution, which serves as a useful prior for sources whose
chirp masses are not so well determined.}
   \label{fig:chirp}
\end{figure}

Figure~\ref{fig:chirp} shows the chirp mass distribution for binaries in our simulated galaxy.  We use this
distribution to construct a hyper-prior on the chirp mass, approximated by the following distribution:
\begin{equation}\label{ChirpDist}
p({\mathcal{M}}) =  \frac{C}{\left(\frac{\mathcal{M}}{\mathcal{M}_0}\right)^{-a} + \frac{a}{b}\left(\frac{\mathcal{M}}{\mathcal{M}_0}\right)^b}
\end{equation}
where $\mathcal{M}_0, a$ and $b$ are hyper-parameters in our model.  $C$ is the normalization constant which can be calculated analytically and is given by:
\begin{equation}\label{ChirpNorm}
C =  \mathcal{M}_0 \pi \frac{b^{\frac{a+1}{a+b}}a^{-\frac{a+1}{a+b}}}{(a+b)\sin{\frac{\pi(b-1)}{a+b}}}
\end{equation}
$\mathcal{M}_0$ is the mode of the distribution. The hyper-parameters $a$ and $b$ determine the width of the distribution, which can be seen by calculating the full width at half maximum (FWHM).  It is given by:
\begin{equation}\label{ChirpFWHM}
\text{FWHM} \simeq \mathcal{M}_0 \left(\left[2(b/a+1)\right]^{1/b} - \left[2(a/b+1)\right]^{-1/a}\right)
\end{equation}

We further assume that the orbital evolution is due only to the emission of gravitational waves, and is thus
adequately described by ($\ref{fdot}$).  In principle
one would want to be more careful and consider tidal effects and mass transfer~\cite{Stroeer:2009uy} as
possible contributions to $\dot{f}$.  However, it is expected that the high frequency sources we are focusing
on will be mostly detached white dwarf binaries where tidal or mass transfer effects are unlikely to
be significant~\cite{Willems:2009xk}.

\section{Results}

We are able to efficiently calculate the full likelihood for each source (eq.~\ref{eq:FullLikelihood}) using the
fast waveform generator developed by Cornish and Littenberg~\cite{Cornish:2007if}.  The following results are
all derived from simulations using the full likelihood.  
Using the same MCMC approach from our toy models, we sample the posterior and get PDFs for 
source and model parameters simultaneously.  We check for convergence by starting the
chains at different locations in the prior volume and find that regardless of starting location, the chains converge to the same PDFs.
  
\begin{figure}[htbp]
   \centering
   \includegraphics[width=3.4in,angle=0] {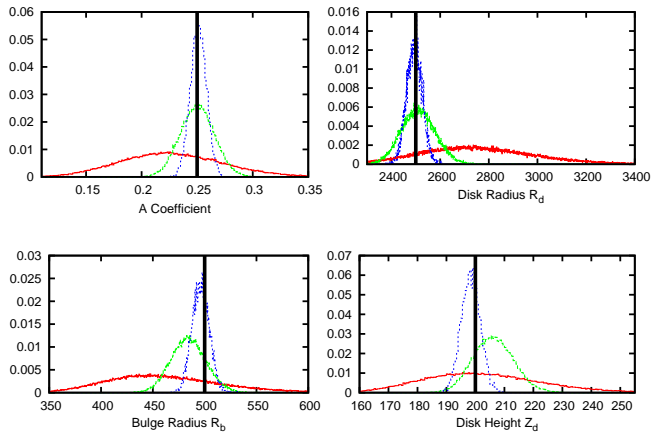} 
   \caption{PDFs for the four galaxy model hyper-parameters.  The red is for a simulation using 100 binaries, the green 1000 binaries,
   and the blue 5000 binaries.  The black lines show the true values of the distribution from which the binaries were drawn.}
   \label{fig:hyperparameters}
\end{figure}

\begin{figure}[htbp]
   \centering
   \includegraphics[width=3.4in,angle=0] {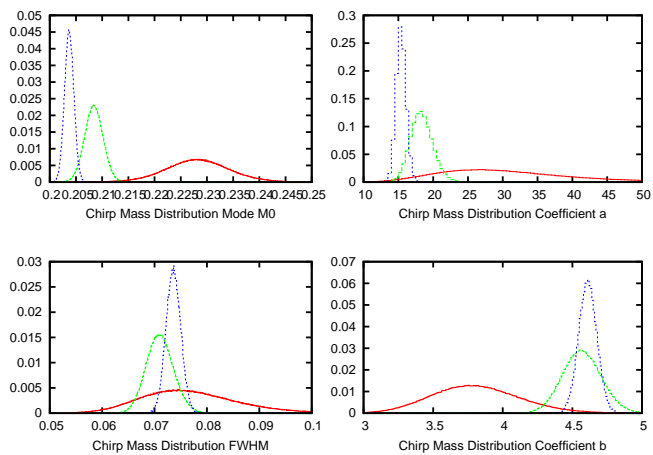} 
   \caption{PDFs for the three chirp mass model hyper-parameters and the FWHM of the distribution.  The red is for a simulation using 100 
   binaries, the green 1000 binaries, and the blue 5000 binaries.}
   \label{fig:chirpparameters}
\end{figure}

Our procedure successfully recovers the correct chirp mass distribution, as shown in Figure~\ref{fig:chirp} and is able to meaningfully constrain the parameters of the galaxy distribution and chirp mass distribution models, with PDFs shown in Figure~\ref{fig:hyperparameters} and Figure~\ref{fig:chirpparameters} respectively.

We ran simulations with 100, 1000, and 5000 binaries to show how the constraints on the galaxy hyper-parameters
improved as we include more sources (for comparison, eLISA is expected to detect between 3500-4100
white dwarf binaries during a 2-year mission lifetime~\cite{AmaroSeoane:2012je}).  The chains ran for
1 million, 500k, and 100k iterations respectively.  Even for a relatively modest number of detections
we begin to get meaningful measurements on the population model of white dwarf binary systems. 
The more binaries we use in our analysis the tighter our constraints on the hyper-parameters.  

\begin{table}[ht]
\centering
\begin{tabular}{c | c c | c c | c c}
\hline\hline
\multicolumn{1}{c|}{} &  \multicolumn{2}{c|}{100} & \multicolumn{2}{c|}{1000} & \multicolumn{2}{c}{5000} \\  
\multicolumn{1}{c|}{Parameter}  & \multicolumn{1}{ c }{MAP} & \multicolumn{1}{ c|}{$\sigma$} 
                                                          & \multicolumn{1}{ c }{MAP} & \multicolumn{1}{ c|}{$\sigma$} 
                                                          & \multicolumn{1}{ c }{MAP} & \multicolumn{1}{ c}{$\sigma$} \\ 
\hline
A           & 0.262 & 0.047  & 0.226 & 0.0157 & 0.249 & 0.0074 \\
Rb (pc) & 440    & 58.9    & 490    & 17.1      & 480    & 8.38 \\
Rd (pc) & 2465  & 237.5 & 2584  & 70.2     & 2461   & 32.4 \\
Zd (pc)  & 193    & 20.8   & 201     & 7.02     & 195     & 3.25 \\ [1ex] 
\hline
$\mathcal{M}_0$ & 0.226 & 0.0063 & 0.208 & 0.0018 & 0.205 & 0.00088 \\
FWHM               & 0.07 & 0.0094 & 0.071& 0.0026 & 0.076 & 0.0014 \\ [1ex] 
\hline
\end{tabular}
\caption{MAP values and variances for the galaxy hyper-parameters when using 100, 1000 and 5000
galactic binaries in the analysis. The simulated values were $A=0.25$,
$R_b=500$pc, $R_d=2500$pc, and $Z_d=200$pc.}
\label{table1}
\end{table}

Table 1 lists the recovered MAP values
and the variance of the marginalized posterior distribution function for each hyper-parameter. Gravitational wave observations
would be very competitive with existing electromagnetic observations in constraining the shape of the
galaxy ~\cite{McMillan:2009yr, Juric:2005zr}.  Making direct comparisons between our results to those in the literature is complicated, as the actual values of the bulge and disk radii are very model dependent.  For example, Juric uses a model
where the galaxy is comprised of both a thin and thick disk.  With GW data in hand, this comparison could easily be made by trivially substituting the density profile used here.  

What matters for this proof-of-principal study is how well the parameters can be constrained.  In the models of Juric et al constraints for the disk radii are around 20\%.  We find similar accuracy when using a pessimistic population of 100 systems.  Adopting a source catalog that is more consistent with theoretical predictions, we find constraints for the disk parameters as low as 1.5\% -- a substantial improvement over the state-of-the-art.

\subsection{Approximating the Likelihood}\label{approx}

\begin{figure}[htbp]
   \centering
   \includegraphics[width=3.4in,angle=0] {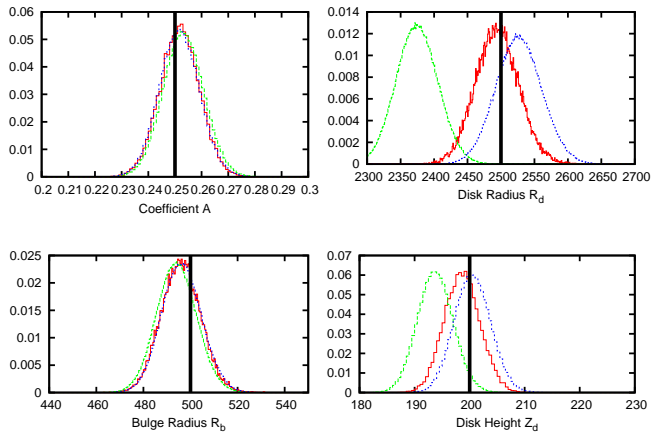} 
   \caption{PDFs from a simulation using 5000 binaries for the four galaxy model hyper-parameters for the full likelihood (red), 
   a Fisher approximation in $d$ (green), and a Fisher approximation in $1/d$ (blue).}
   \label{fig:CompareL}
\end{figure}

While we happen to have a very efficient method for computing the full likelihood for galactic binaries, this
is not always the case. For other signal types the full likelihood can be very expensive to compute, posing
problems if we wish to do extensive studies
of many astrophysical models or detector configurations. For such exploratory studies it is preferable to use
the Fisher Information Matrix approximation to the likelihood of (\ref{fish}). However, as we saw with the toy
model in \S\ref{toy2}, this can lead to biases in the recovered parameters.
The Fisher matrix $\Gamma_{ij}$ is not a
coordinate invariant quantity, and we can at least partially correct the bias by reparameterizing our likelihood.
Just as in \S\ref{toy2}, instead of using the distance $d$ as a variable, we can instead use $1/d$, which provides a much better
approximation to the full likelihood.  We test these short-cuts by redoing the analysis of the galactic population using the Fisher matrix approximation to the likelihood (both with $d$ and $1/d$ as parameters) and compare to the results from the previous analysis using the full likelihood.
Figure~\ref{fig:CompareL} shows PDFs for the galaxy hyper-parameters
using the three different methods for computing $p(d|\vec{\lambda})$ with the full sample of 5000 binaries.
  
We find that the approximation using $1/d$ matches the full likelihood better than the likelihood parameterized with $d$, however there
are additional discrepancies due to non-quadratic terms in the sky location $\{\theta,\phi\}$ that we have not accounted
for. The dependence of the waveform on $\{\theta,\phi\}$ is more complicated than the distance, and is not so easily
corrected by a simple reparameterization. The approximation could be improved by carrying the expansion of the
likelihood beyond second order, however this is computationally expensive and can be numerically unstable.

\begin{figure}[htbp]
   \centering
   \includegraphics[width=3.4in,angle=0] {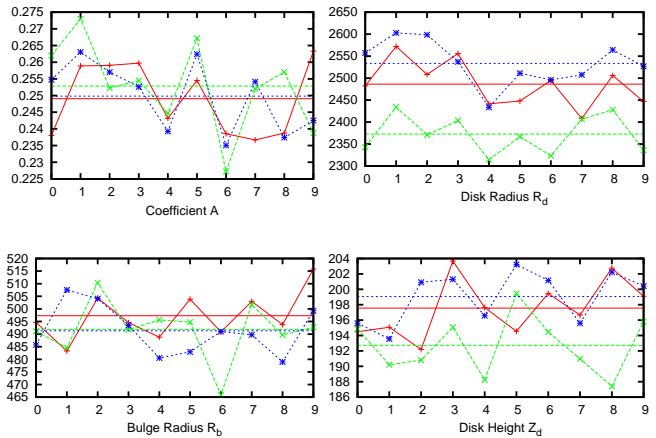} 
   \caption{MAP values and corresponding averages from a simulations using 5000 binaries for the four
galaxy model hyper-parameters for the full likelihood (red), a Fisher Matrix approximation parameterized with $d$ (green),
and a Fisher Matrix approximation using $1/d$ (blue).}
   \label{fig:CompareMAPs}
\end{figure}

If we analyze several realizations of the galaxy using the three different likelihood functions and average the results, we find the biases are persistent for the approximate methods. Figure~\ref{fig:CompareMAPs} shows the MAP values and the average of the MAP values for 10 realizations of
our fiducial galaxy model.  The biases in the recovered disk radius and disk height are particularly
pronounced when using the Fisher Matrix approximation to the likelihood parameterized with $d$. 

\section{Conclusion}\label{concl}
We have demonstrated a general Hierarchical Bayesian method capable of constraining the model parameters for a
population of sources.  In the particular case of white dwarf binaries in the Milky Way, we can
constrain the spatial distribution of the galaxy to levels better than current electromagnetic observations using the anticipated number of systems detectable by space-based gravitational wave detectors.  
Even if the currently held event rates for white dwarf binaries turn out to be optimistic by more than an order of magnitude, the constraints possible with a gravitational wave detector are comparable to our current estimates of the Milky Way's shape.  

When the data from a space-borne detector has been collected, the resolvable white dwarf binaries will be regressed from the data, leaving behind a confusion-limited foreground which will significantly contribute to the overall power in the data around $\sim 1$ mHz.
Measuring the overall shape of the galaxy as demonstrated here will provide additional means to characterize the level of the confusion noise.  As we will show in an upcoming paper, we can then use the detailed understanding of the foreground signal to detect a stochastic gravitational wave background at levels well below the confusion noise.

Analyzing simulated data with the full likelihood is computationally taxing and, when performing a large suite of 
such studies, could prove to be prohibitive.  To mitigate the cost of such analyses, we test a much faster approach (approximately 50 times faster), using 
the Fisher matrix approximation to the likelihood.  We find the results are significantly less biased by the Fisher approximation when 
using $1/d$ as the parameter that encodes the distance to the source.  This simple adjustment gives adequately reliable results in significantly less time than the brute-force calculation, and will provide an additional, useful, metric to gauge the relative merits of proposed 
space-based gravitational wave missions.

\section{Acknowledgments}
NJC and MA were supported by NASA grant NNX07AJ61G.  TBL was supported by NASA Grant 08-ATFP08-0126.

\bibliography{astromodel.bib}

\begin{thebibliography}{43}
\expandafter\ifx\csname natexlab\endcsname\relax\def\natexlab#1{#1}\fi
\expandafter\ifx\csname bibnamefont\endcsname\relax
  \def\bibnamefont#1{#1}\fi
\expandafter\ifx\csname bibfnamefont\endcsname\relax
  \def\bibfnamefont#1{#1}\fi
\expandafter\ifx\csname citenamefont\endcsname\relax
  \def\citenamefont#1{#1}\fi
\expandafter\ifx\csname url\endcsname\relax
  \def\url#1{\texttt{#1}}\fi
\expandafter\ifx\csname urlprefix\endcsname\relax\def\urlprefix{URL }\fi
\providecommand{\bibinfo}[2]{#2}
\providecommand{\eprint}[2][]{\url{#2}}

\bibitem[{\citenamefont{{Stebbins}}(2009)}]{2009CQGra..26i4014S}
\bibinfo{author}{\bibfnamefont{R.~T.} \bibnamefont{{Stebbins}}},
  \bibinfo{journal}{Classical and Quantum Gravity}
  \textbf{\bibinfo{volume}{26}}, \bibinfo{pages}{094014}
  (\bibinfo{year}{2009}), \eprint{0904.1029}.

\bibitem[{\citenamefont{Gair et~al.}(2010)\citenamefont{Gair, Tang, and
  Volonteri}}]{Gair:2010yu}
\bibinfo{author}{\bibfnamefont{J.~R.} \bibnamefont{Gair}},
  \bibinfo{author}{\bibfnamefont{C.}~\bibnamefont{Tang}}, \bibnamefont{and}
  \bibinfo{author}{\bibfnamefont{M.}~\bibnamefont{Volonteri}},
  \bibinfo{journal}{Phys.Rev.} \textbf{\bibinfo{volume}{D81}},
  \bibinfo{pages}{104014} (\bibinfo{year}{2010}), \eprint{1004.1921}.

\bibitem[{\citenamefont{Gair et~al.}(2011)\citenamefont{Gair, Sesana, Berti,
  and Volonteri}}]{Gair:2010bx}
\bibinfo{author}{\bibfnamefont{J.~R.} \bibnamefont{Gair}},
  \bibinfo{author}{\bibfnamefont{A.}~\bibnamefont{Sesana}},
  \bibinfo{author}{\bibfnamefont{E.}~\bibnamefont{Berti}}, \bibnamefont{and}
  \bibinfo{author}{\bibfnamefont{M.}~\bibnamefont{Volonteri}},
  \bibinfo{journal}{Class.Quant.Grav.} \textbf{\bibinfo{volume}{28}},
  \bibinfo{pages}{094018} (\bibinfo{year}{2011}), \eprint{1009.6172}.

\bibitem[{\citenamefont{Sesana et~al.}(2011)\citenamefont{Sesana, Gair, Berti,
  and Volonteri}}]{Sesana:2010wy}
\bibinfo{author}{\bibfnamefont{A.}~\bibnamefont{Sesana}},
  \bibinfo{author}{\bibfnamefont{J.}~\bibnamefont{Gair}},
  \bibinfo{author}{\bibfnamefont{E.}~\bibnamefont{Berti}}, \bibnamefont{and}
  \bibinfo{author}{\bibfnamefont{M.}~\bibnamefont{Volonteri}},
  \bibinfo{journal}{Phys.Rev.} \textbf{\bibinfo{volume}{D83}},
  \bibinfo{pages}{044036} (\bibinfo{year}{2011}), \eprint{1011.5893}.

\bibitem[{\citenamefont{Amaro-Seoane et~al.}(2012)\citenamefont{Amaro-Seoane,
  Aoudia, Babak, Binetruy, Berti et~al.}}]{AmaroSeoane:2012je}
\bibinfo{author}{\bibfnamefont{P.}~\bibnamefont{Amaro-Seoane}},
  \bibinfo{author}{\bibfnamefont{S.}~\bibnamefont{Aoudia}},
  \bibinfo{author}{\bibfnamefont{S.}~\bibnamefont{Babak}},
  \bibinfo{author}{\bibfnamefont{P.}~\bibnamefont{Binetruy}},
  \bibinfo{author}{\bibfnamefont{E.}~\bibnamefont{Berti}},
  \bibnamefont{et~al.}, \bibinfo{journal}{Class.Quant.Grav.}
  \textbf{\bibinfo{volume}{29}}, \bibinfo{pages}{124016}
  (\bibinfo{year}{2012}), \eprint{1202.0839}.

\bibitem[{\citenamefont{Plowman et~al.}(2009)\citenamefont{Plowman, Jacobs,
  Hellings, Larson, and Tsuruta}}]{Plowman:2009rp}
\bibinfo{author}{\bibfnamefont{J.~E.} \bibnamefont{Plowman}},
  \bibinfo{author}{\bibfnamefont{D.~C.} \bibnamefont{Jacobs}},
  \bibinfo{author}{\bibfnamefont{R.~W.} \bibnamefont{Hellings}},
  \bibinfo{author}{\bibfnamefont{S.~L.} \bibnamefont{Larson}},
  \bibnamefont{and} \bibinfo{author}{\bibfnamefont{S.}~\bibnamefont{Tsuruta}}
  (\bibinfo{year}{2009}), \eprint{0903.2059}.

\bibitem[{\citenamefont{Plowman et~al.}(2010)\citenamefont{Plowman, Hellings,
  and Tsuruta}}]{Plowman:2010fc}
\bibinfo{author}{\bibfnamefont{J.~E.} \bibnamefont{Plowman}},
  \bibinfo{author}{\bibfnamefont{R.~W.} \bibnamefont{Hellings}},
  \bibnamefont{and} \bibinfo{author}{\bibfnamefont{S.}~\bibnamefont{Tsuruta}}
  (\bibinfo{year}{2010}), \eprint{1009.0765}.

\bibitem[{\citenamefont{Mandel}(2010)}]{Mandel:2009pc}
\bibinfo{author}{\bibfnamefont{I.}~\bibnamefont{Mandel}},
  \bibinfo{journal}{Phys.Rev.} \textbf{\bibinfo{volume}{D81}},
  \bibinfo{pages}{084029} (\bibinfo{year}{2010}), \eprint{0912.5531}.

\bibitem[{\citenamefont{O'Shaughnessy}(2012)}]{O'Shaughnessy:2012zc}
\bibinfo{author}{\bibfnamefont{R.}~\bibnamefont{O'Shaughnessy}}
  (\bibinfo{year}{2012}), \eprint{1204.3117}.

\bibitem[{\citenamefont{{Mandel} et~al.}(2009)\citenamefont{{Mandel},
  {Wood-Vasey}, {Friedman}, and {Kirshner}}}]{2009ApJ...704..629M}
\bibinfo{author}{\bibfnamefont{K.~S.} \bibnamefont{{Mandel}}},
  \bibinfo{author}{\bibfnamefont{W.~M.} \bibnamefont{{Wood-Vasey}}},
  \bibinfo{author}{\bibfnamefont{A.~S.} \bibnamefont{{Friedman}}},
  \bibnamefont{and} \bibinfo{author}{\bibfnamefont{R.~P.}
  \bibnamefont{{Kirshner}}}, \bibinfo{journal}{\apj}
  \textbf{\bibinfo{volume}{704}}, \bibinfo{pages}{629} (\bibinfo{year}{2009}),
  \eprint{0908.0536}.

\bibitem[{\citenamefont{{Soiaporn} et~al.}(2012)\citenamefont{{Soiaporn},
  {Chernoff}, {Loredo}, {Ruppert}, and {Wasserman}}}]{2012arXiv1206.3540S}
\bibinfo{author}{\bibfnamefont{K.}~\bibnamefont{{Soiaporn}}},
  \bibinfo{author}{\bibfnamefont{D.}~\bibnamefont{{Chernoff}}},
  \bibinfo{author}{\bibfnamefont{T.}~\bibnamefont{{Loredo}}},
  \bibinfo{author}{\bibfnamefont{D.}~\bibnamefont{{Ruppert}}},
  \bibnamefont{and}
  \bibinfo{author}{\bibfnamefont{I.}~\bibnamefont{{Wasserman}}},
  \bibinfo{journal}{ArXiv e-prints}  (\bibinfo{year}{2012}),
  \eprint{1206.3540}.

\bibitem[{\citenamefont{{Loredo}}(2012)}]{2012arXiv1208.3036L}
\bibinfo{author}{\bibfnamefont{T.~J.} \bibnamefont{{Loredo}}},
  \bibinfo{journal}{ArXiv e-prints}  (\bibinfo{year}{2012}),
  \eprint{1208.3036}.

\bibitem[{\citenamefont{Messenger and Veitch}(2012)}]{Messenger:2012jy}
\bibinfo{author}{\bibfnamefont{C.}~\bibnamefont{Messenger}} \bibnamefont{and}
  \bibinfo{author}{\bibfnamefont{J.}~\bibnamefont{Veitch}}
  (\bibinfo{year}{2012}), \eprint{1206.3461}.

\bibitem[{\citenamefont{Schutz}(2011)}]{Schutz:2011tw}
\bibinfo{author}{\bibfnamefont{B.~F.} \bibnamefont{Schutz}},
  \bibinfo{journal}{Class.Quant.Grav.} \textbf{\bibinfo{volume}{28}},
  \bibinfo{pages}{125023} (\bibinfo{year}{2011}), \eprint{1102.5421}.

\bibitem[{\citenamefont{Crowder and Cornish}(2007)}]{Crowder:2006eu}
\bibinfo{author}{\bibfnamefont{J.}~\bibnamefont{Crowder}} \bibnamefont{and}
  \bibinfo{author}{\bibfnamefont{N.}~\bibnamefont{Cornish}},
  \bibinfo{journal}{Phys.Rev.} \textbf{\bibinfo{volume}{D75}},
  \bibinfo{pages}{043008} (\bibinfo{year}{2007}), \eprint{astro-ph/0611546}.

\bibitem[{\citenamefont{Nissanke et~al.}(2012)\citenamefont{Nissanke,
  Vallisneri, Nelemans, and Prince}}]{Nissanke:2012eh}
\bibinfo{author}{\bibfnamefont{S.}~\bibnamefont{Nissanke}},
  \bibinfo{author}{\bibfnamefont{M.}~\bibnamefont{Vallisneri}},
  \bibinfo{author}{\bibfnamefont{G.}~\bibnamefont{Nelemans}}, \bibnamefont{and}
  \bibinfo{author}{\bibfnamefont{T.~A.} \bibnamefont{Prince}}
  (\bibinfo{year}{2012}), \eprint{1201.4613}.

\bibitem[{\citenamefont{Timpano et~al.}(2006)\citenamefont{Timpano, Rubbo, and
  Cornish}}]{Timpano:2005gm}
\bibinfo{author}{\bibfnamefont{S.~E.} \bibnamefont{Timpano}},
  \bibinfo{author}{\bibfnamefont{L.~J.} \bibnamefont{Rubbo}}, \bibnamefont{and}
  \bibinfo{author}{\bibfnamefont{N.~J.} \bibnamefont{Cornish}},
  \bibinfo{journal}{Phys.Rev.} \textbf{\bibinfo{volume}{D73}},
  \bibinfo{pages}{122001} (\bibinfo{year}{2006}), \eprint{gr-qc/0504071}.

\bibitem[{\citenamefont{Littenberg}(2011)}]{Littenberg:2011zg}
\bibinfo{author}{\bibfnamefont{T.~B.} \bibnamefont{Littenberg}},
  \bibinfo{journal}{Phys.Rev.} \textbf{\bibinfo{volume}{D84}},
  \bibinfo{pages}{063009} (\bibinfo{year}{2011}), \eprint{1106.6355}.

\bibitem[{\citenamefont{{Adams} and {Cornish}}(2010)}]{2010PhRvD..82b2002A}
\bibinfo{author}{\bibfnamefont{M.~R.} \bibnamefont{{Adams}}} \bibnamefont{and}
  \bibinfo{author}{\bibfnamefont{N.~J.} \bibnamefont{{Cornish}}},
  \bibinfo{journal}{\prd} \textbf{\bibinfo{volume}{82}}, \bibinfo{eid}{022002}
  (\bibinfo{year}{2010}), \eprint{1002.1291}.

\bibitem[{\citenamefont{Good}(1965)}]{Good:1965}
\bibinfo{author}{\bibfnamefont{I.~J.} \bibnamefont{Good}},
  \emph{\bibinfo{title}{{The Estimation of Probabilities: An Essay on Modern
  Bayesian Methods}}} (\bibinfo{publisher}{MIT Press, Cambridge, Mass.},
  \bibinfo{year}{1965}).

\bibitem[{\citenamefont{Lindley and Smith}(1972)}]{1972_LindleySmith}
\bibinfo{author}{\bibfnamefont{D.~V.} \bibnamefont{Lindley}} \bibnamefont{and}
  \bibinfo{author}{\bibfnamefont{A.~F.~M.} \bibnamefont{Smith}},
  \bibinfo{journal}{Journal of the Royal Statistical Society. Series B
  (Methodological)} \textbf{\bibinfo{volume}{34}}, \bibinfo{pages}{1}
  (\bibinfo{year}{1972}), ISSN \bibinfo{issn}{00359246},
  \urlprefix\url{http://dx.doi.org/10.2307/2985048}.

\bibitem[{\citenamefont{Morris and Normand}(1992)}]{Morris:1992}
\bibinfo{author}{\bibfnamefont{C.~N.} \bibnamefont{Morris}} \bibnamefont{and}
  \bibinfo{author}{\bibfnamefont{S.~L.} \bibnamefont{Normand}}, in
  \emph{\bibinfo{booktitle}{Bayesian Statistics 4}}, edited by
  \bibinfo{editor}{\bibfnamefont{A.~P.~D.} \bibnamefont{J.~M.~Bernardo},
  \bibfnamefont{J.~O.~Berger}} \bibnamefont{and}
  \bibinfo{editor}{\bibfnamefont{A.~F.~M.} \bibnamefont{Smith}}
  (\bibinfo{publisher}{Oxford Univeristy Press, Oxford}, \bibinfo{year}{1992}),
  pp. \bibinfo{pages}{321--344}.

\bibitem[{\citenamefont{MacKay}(1999)}]{MacKay94}
\bibinfo{author}{\bibfnamefont{D.~J.~C.} \bibnamefont{MacKay}},
  \bibinfo{journal}{Neural Computation} \textbf{\bibinfo{volume}{11}},
  \bibinfo{pages}{1035} (\bibinfo{year}{1999}).

\bibitem[{\citenamefont{Casella}(1985)}]{Casella:1985}
\bibinfo{author}{\bibfnamefont{G.}~\bibnamefont{Casella}},
  \bibinfo{journal}{The American Statistician} \textbf{\bibinfo{volume}{39}},
  \bibinfo{pages}{83} (\bibinfo{year}{1985}).

\bibitem[{\citenamefont{Carlin}(2000)}]{Carlin:2000}
\bibinfo{author}{\bibfnamefont{T.~A.} \bibnamefont{Carlin},
  \bibfnamefont{Bradley P.;~Louis}}, \emph{\bibinfo{title}{Bayes and Empirical
  Bayes Methods for Data Analysis}} (\bibinfo{publisher}{Chapman \& Hall/CRC.},
  \bibinfo{year}{2000}), \bibinfo{edition}{2nd} ed.

\bibitem[{\citenamefont{Cornish and Littenberg}(2007)}]{Cornish:2007if}
\bibinfo{author}{\bibfnamefont{N.~J.} \bibnamefont{Cornish}} \bibnamefont{and}
  \bibinfo{author}{\bibfnamefont{T.~B.} \bibnamefont{Littenberg}},
  \bibinfo{journal}{Phys.Rev.} \textbf{\bibinfo{volume}{D76}},
  \bibinfo{pages}{083006} (\bibinfo{year}{2007}), \eprint{0704.1808}.

\bibitem[{\citenamefont{Littenberg and Cornish}(2010)}]{Littenberg:2010gf}
\bibinfo{author}{\bibfnamefont{T.~B.} \bibnamefont{Littenberg}}
  \bibnamefont{and} \bibinfo{author}{\bibfnamefont{N.~J.}
  \bibnamefont{Cornish}}, \bibinfo{journal}{Phys.Rev.}
  \textbf{\bibinfo{volume}{D82}}, \bibinfo{pages}{103007}
  (\bibinfo{year}{2010}), \eprint{1008.1577}.

\bibitem[{\citenamefont{Green et~al.}(2003)\citenamefont{Green, Hjort, and
  Richardson}}]{green_highly_2003}
\bibinfo{author}{\bibfnamefont{P.~J.} \bibnamefont{Green}},
  \bibinfo{author}{\bibfnamefont{N.~L.} \bibnamefont{Hjort}}, \bibnamefont{and}
  \bibinfo{author}{\bibfnamefont{S.}~\bibnamefont{Richardson}},
  \emph{\bibinfo{title}{Highly Structured Stochastic Systems}}
  (\bibinfo{publisher}{Oxford University Press}, \bibinfo{year}{2003}), ISBN
  \bibinfo{isbn}{9780198510550}.

\bibitem[{\citenamefont{Cornish and Larson}(2003)}]{Cornish:2003vj}
\bibinfo{author}{\bibfnamefont{N.~J.} \bibnamefont{Cornish}} \bibnamefont{and}
  \bibinfo{author}{\bibfnamefont{S.~L.} \bibnamefont{Larson}},
  \bibinfo{journal}{Phys.Rev.} \textbf{\bibinfo{volume}{D67}},
  \bibinfo{pages}{103001} (\bibinfo{year}{2003}), \eprint{astro-ph/0301548}.

\bibitem[{\citenamefont{Vecchio and Wickham}(2004)}]{Vecchio:2004ec}
\bibinfo{author}{\bibfnamefont{A.}~\bibnamefont{Vecchio}} \bibnamefont{and}
  \bibinfo{author}{\bibfnamefont{E.~D.} \bibnamefont{Wickham}},
  \bibinfo{journal}{Phys.Rev.} \textbf{\bibinfo{volume}{D70}},
  \bibinfo{pages}{082002} (\bibinfo{year}{2004}), \eprint{gr-qc/0406039}.

\bibitem[{\citenamefont{B\l{}aut et~al.}(2010)\citenamefont{B\l{}aut, Babak,
  and Kr\'olak}}]{babak:WDs}
\bibinfo{author}{\bibfnamefont{A.}~\bibnamefont{B\l{}aut}},
  \bibinfo{author}{\bibfnamefont{S.}~\bibnamefont{Babak}}, \bibnamefont{and}
  \bibinfo{author}{\bibfnamefont{A.}~\bibnamefont{Kr\'olak}},
  \bibinfo{journal}{Phys. Rev. D} \textbf{\bibinfo{volume}{81}},
  \bibinfo{pages}{063008} (\bibinfo{year}{2010}),
  \urlprefix\url{http://link.aps.org/doi/10.1103/PhysRevD.81.063008}.

\bibitem[{\citenamefont{et~al.}(2009)}]{LISAwhitepaper}
\bibinfo{author}{\bibfnamefont{R.~S.} \bibnamefont{et~al.}},
  \emph{\bibinfo{title}{Laser interferometer space antenna (lisa) a response to
  the astro2010 rfi for the particle astrophysics and gravitation panel}},
  \bibinfo{howpublished}{White paper, NASA} (\bibinfo{year}{2009}),
  \bibinfo{note}{available online},
  \urlprefix\url{http://lisa.nasa.gov/documentation.html}.

\bibitem[{\citenamefont{Schutz}(1986)}]{Schutz:1986gp}
\bibinfo{author}{\bibfnamefont{B.~F.} \bibnamefont{Schutz}},
  \bibinfo{journal}{Nature} \textbf{\bibinfo{volume}{323}},
  \bibinfo{pages}{310} (\bibinfo{year}{1986}).

\bibitem[{\citenamefont{Takahashi and Seto}(2002)}]{Takahashi:2002ky}
\bibinfo{author}{\bibfnamefont{R.}~\bibnamefont{Takahashi}} \bibnamefont{and}
  \bibinfo{author}{\bibfnamefont{N.}~\bibnamefont{Seto}},
  \bibinfo{journal}{Astrophys.J.} \textbf{\bibinfo{volume}{575}},
  \bibinfo{pages}{1030} (\bibinfo{year}{2002}), \eprint{astro-ph/0204487}.

\bibitem[{\citenamefont{Nelemans et~al.}(2004)\citenamefont{Nelemans,
  Yungelson, and Portegies~Zwart}}]{Nelemans:2003ha}
\bibinfo{author}{\bibfnamefont{G.}~\bibnamefont{Nelemans}},
  \bibinfo{author}{\bibfnamefont{L.}~\bibnamefont{Yungelson}},
  \bibnamefont{and}
  \bibinfo{author}{\bibfnamefont{S.}~\bibnamefont{Portegies~Zwart}},
  \bibinfo{journal}{Mon.Not.Roy.Astron.Soc.} \textbf{\bibinfo{volume}{349}},
  \bibinfo{pages}{181} (\bibinfo{year}{2004}), \eprint{astro-ph/0312193}.

\bibitem[{\citenamefont{Nelemans
  et~al.}(2001{\natexlab{a}})\citenamefont{Nelemans, Yungelson,
  Portegies~Zwart, and Verbunt}}]{Nelemans:2000es}
\bibinfo{author}{\bibfnamefont{G.}~\bibnamefont{Nelemans}},
  \bibinfo{author}{\bibfnamefont{L.~R.} \bibnamefont{Yungelson}},
  \bibinfo{author}{\bibfnamefont{S.~F.} \bibnamefont{Portegies~Zwart}},
  \bibnamefont{and} \bibinfo{author}{\bibfnamefont{F.}~\bibnamefont{Verbunt}},
  \bibinfo{journal}{Astron.Astrophys.} \textbf{\bibinfo{volume}{365}},
  \bibinfo{pages}{491} (\bibinfo{year}{2001}{\natexlab{a}}),
  \eprint{astro-ph/0010457}.

\bibitem[{\citenamefont{Nelemans
  et~al.}(2001{\natexlab{b}})\citenamefont{Nelemans, Portegies~Zwart, Verbunt,
  and Yungelson}}]{Nelemans:2001nr}
\bibinfo{author}{\bibfnamefont{G.}~\bibnamefont{Nelemans}},
  \bibinfo{author}{\bibfnamefont{S.~F.} \bibnamefont{Portegies~Zwart}},
  \bibinfo{author}{\bibfnamefont{F.}~\bibnamefont{Verbunt}}, \bibnamefont{and}
  \bibinfo{author}{\bibfnamefont{L.}~\bibnamefont{Yungelson}},
  \bibinfo{journal}{Astron.Astrophys.} \textbf{\bibinfo{volume}{368}},
  \bibinfo{pages}{939} (\bibinfo{year}{2001}{\natexlab{b}}),
  \eprint{astro-ph/0101123}.

\bibitem[{\citenamefont{Nelemans
  et~al.}(2001{\natexlab{c}})\citenamefont{Nelemans, Yungelson, and
  Portegies~Zwart}}]{Nelemans:2001hp}
\bibinfo{author}{\bibfnamefont{G.}~\bibnamefont{Nelemans}},
  \bibinfo{author}{\bibfnamefont{L.}~\bibnamefont{Yungelson}},
  \bibnamefont{and} \bibinfo{author}{\bibfnamefont{S.~F.}
  \bibnamefont{Portegies~Zwart}}, \bibinfo{journal}{Astron.Astrophys.}
  \textbf{\bibinfo{volume}{375}}, \bibinfo{pages}{890}
  (\bibinfo{year}{2001}{\natexlab{c}}), \eprint{astro-ph/0105221}.

\bibitem[{\citenamefont{Arnaud et~al.}(2007)\citenamefont{Arnaud, Babak, Baker,
  Benacquista, Cornish et~al.}}]{Arnaud:2007jy}
\bibinfo{author}{\bibfnamefont{K.}~\bibnamefont{Arnaud}},
  \bibinfo{author}{\bibfnamefont{S.}~\bibnamefont{Babak}},
  \bibinfo{author}{\bibfnamefont{J.}~\bibnamefont{Baker}},
  \bibinfo{author}{\bibfnamefont{M.}~\bibnamefont{Benacquista}},
  \bibinfo{author}{\bibfnamefont{N.}~\bibnamefont{Cornish}},
  \bibnamefont{et~al.}, \bibinfo{journal}{Class.Quant.Grav.}
  \textbf{\bibinfo{volume}{24}}, \bibinfo{pages}{S551} (\bibinfo{year}{2007}),
  \eprint{gr-qc/0701170}.

\bibitem[{\citenamefont{Stroeer and Nelemans}(2009)}]{Stroeer:2009uy}
\bibinfo{author}{\bibfnamefont{A.}~\bibnamefont{Stroeer}} \bibnamefont{and}
  \bibinfo{author}{\bibfnamefont{G.}~\bibnamefont{Nelemans}},
  \bibinfo{journal}{Mon.Not.Roy.Astron.Soc.Lett.}  (\bibinfo{year}{2009}),
  \eprint{0909.1796}.

\bibitem[{\citenamefont{Willems et~al.}(2010)\citenamefont{Willems, Deloye, and
  Kalogera}}]{Willems:2009xk}
\bibinfo{author}{\bibfnamefont{B.}~\bibnamefont{Willems}},
  \bibinfo{author}{\bibfnamefont{C.}~\bibnamefont{Deloye}}, \bibnamefont{and}
  \bibinfo{author}{\bibfnamefont{V.}~\bibnamefont{Kalogera}},
  \bibinfo{journal}{Astrophys.J.} \textbf{\bibinfo{volume}{713}},
  \bibinfo{pages}{239} (\bibinfo{year}{2010}), \eprint{0904.1953}.

\bibitem[{\citenamefont{McMillan and Binney}(2009)}]{McMillan:2009yr}
\bibinfo{author}{\bibfnamefont{P.~J.} \bibnamefont{McMillan}} \bibnamefont{and}
  \bibinfo{author}{\bibfnamefont{J.~J.} \bibnamefont{Binney}}
  (\bibinfo{year}{2009}), \eprint{0907.4685}.

\bibitem[{\citenamefont{Juric et~al.}(2008)}]{Juric:2005zr}
\bibinfo{author}{\bibfnamefont{M.}~\bibnamefont{Juric}} \bibnamefont{et~al.}
  (\bibinfo{collaboration}{SDSS Collaboration}),
  \bibinfo{journal}{Astrophys.J.} \textbf{\bibinfo{volume}{673}},
  \bibinfo{pages}{864} (\bibinfo{year}{2008}), \eprint{astro-ph/0510520}.

\end{thebibliography}


\begin{thebibliography}{99} 




\bibitem{lindley}
D.~V.~Lindley and A.~F.~M.~Smith,
J. Roy. Statist. Soc. B{\bf 34}, 1 (1972).



\end{thebibliography}

\end{document}